\documentstyle[12pt]{article}
\def\beq{\begin{equation}}
\def\eeq{\end{equation}}
\def\bea{\begin{eqnarray}}
\def\eea{\end{eqnarray}}

\def\ba{\begin{array}}
\def\ea{\end{array}}

\def\,{\"{U}}
\def\6{\.{I}}

\begin{document}
\baselineskip 0.7cm

\title{\textbf{A Novel SUSY Energy Bound States Treatment of the
Klein-Gordon Equation with \emph{PT}-Supersymmetric and \emph{q}-Deformed \\ Hulth\'{e}n Potential}}

\author{Metin Akta\c{s}$
\thanks{E-mail: maktas@ybu.edu.tr}$\\[0.3cm]
Energy Systems Engineering Department, Engineering and
Natural Sciences \\
Faculty, Ankara Y{\i}ld{\i}r{\i}m Beyaz{\i}t University, 06030, Ankara, Turkey}

\date{\today}

\maketitle

\begin{abstract}

\noindent In this study, we focus on investigating the exact relativistic bound state spectra for supersymmetric, \emph{PT}-supersymmetric and \emph{non}-Hermitian versions of \emph{q}-deformed parameter Hulth\'{e}n potential. The Hamiltonian hierarchy mechanism, namely the \emph{factorization method}, is adopted within the framework of SUSYQM. This algebraic approach is used in solving of the Klein-Gordon equation with the potential cases. The results obtained analytically by executing the straightforward calculations are in consistent forms for certain values of \emph{q}. Achieving the results may have a particular interest for such applications. That is, they can be involved in determining the quantum structural properties of molecules for ro-vibrational states, and optical spectra characteristics of semiconductor devices with regard to the lattice dynamics [62-64]. They are also employed to construct broken or unbroken case of supersymmetric particle model concerning the interaction between the elementary particles [65].\\

\smallskip
\noindent \textbf{PACS numbers:} 03.65.Ca; 03.65.Ge; 03.65.Pm \\[0.3cm]
\noindent \textbf{Keywords:} Klein-Gordon equation, Relativistic bound spectra, Hulth\'{e}n potential, Hamiltonian hierarchy method, Supersymmetry, PT-symmetry, PT-Supersymmetric quantum mechanics, Non-Hermitian operators.
\end{abstract}

\newpage
\section{Introduction}
\noindent The dynamical, spontaneous and meta-stable supersymmetry breaking mechanisms [1-8] arising from the concept of supersymmetry [9-14] are of great importance for quantum mechanical systems involved in a symmetry viewpoint. Recent studies and interests on quantum mechanical systems with supersymmetry as well as on \emph{PT}-symmetric Hamiltonian systems [15-21] have just been in progress. In addition, their significant impacts have just kept going for a long time [22-29]. These type of systems which are Hermitian
require the real and discrete energy eigenvalues. The concept of \emph{PT}-symmetry (\emph{i. e.}, space-time reflectional symmetry) referring to the ($\emph{P}\rightarrow$\emph{parity}) and the ($\emph{T}\rightarrow$\emph{time reversal}) allows us to understand the characteristics of complex potential systems. It is suggested that the \emph{PT}-symmetric but non-Hermitian
Hamiltonians  can be characterized by real spectra for spontaneously
broken \emph{PT}-symmetry [15]. The other crucial concept for a class of these type
Hamiltonians is the pseudo-Hermiticity [30-33]. The pseudo-Hermitian
Hamiltonians imply the real or the complex eigenvalue spectra.\\[0.3cm]
A variety of techniques such as Nikiforov-Uvarov method was performed in solving of the Klein-Gordon (KG) equation within the PT symmetric case [34-36], variational, semi-classical estimates, group theoretical approach, relativistic shifted \textit{l}-expansion and large-\textit{N} expansion [37-50]. Moreover, several applications in the solutions of the KG equation with scalar and vector type potentials (\emph{e. g.}, harmonic oscillator, ring-shaped, exponential types etc.) were introduced by authors [51-61]. Recent engineering applications based on the concepts of supersymmetry and PT-symmetry for optical system devices have also been implemented [62-65].\\[0.3cm]
In this work, we investigate
the exact relativistic bound states of the Klein-Gordon equation for \emph{PT}-supersymmetric
and non-Hermitian Hulth\'{e}n potential cases within the formalism of
SUSYQM [9]. The SUSYQM approach provides an alternative tool to
investigate the potential problems algebraically in relativistic and
non-relativistic quantum mechanics. Several interesting features of SUSYQM are
suggested in this context. It is mainly concerned with both the shape invariance
requirement and the Hamiltonian hierarchy mechanism.\\[0.3cm]
This paper is organized as follows: In Sec. II, hierarchy of a hamiltonian procedure, known as the
\emph{factorization method} is introduced. In Sec. III , we concern with the energy
eigenvalue problem of the Klein-Gordon equation for \emph{q}-deformed
Hulth\'{e}n potential and its \emph{PT}-symmetric, \emph{PT}-supersymmetric and non-Hermitian versions.
Sec. IV deals with the discussion of the results.

\section{Hierarchy of a Hamiltonian: Factorization Method}
\noindent
The main concern of this approach is the reducing of second order ordinary differential equations (ODEs), \emph{e. g.}, Schr\"{o}dinger equation for a given potential to the first order nonlinear Riccati equation involving the superpotential $W(x)$.
This method was first proposed by Darboux and later was
provided in the derivation of exact spectra of hydrogen atom problem
by Schr\"{o}dinger. The SUSYQM algebraic approach allows us to write Hamiltonians by
setting $(\hbar=2m=1)$ [11]

\begin{equation}
\hat{H}_{1}=-\frac{d^{2}}{dx^{2}}+V_{1}(x),\qquad
\hat{H}_{2}=-\frac{d^{2}}{dx^{2}} +V_{2}(x),
\end{equation}\\
\noindent where the superpartner potentials $V_{1,~2}(x)$ in terms of
the superpotential $W(x)$ are given

\begin{equation}
V_{1}(x)=W^{2}-W^{\prime},\qquad V_{2}(x)=W^{2}+W^{\prime}.
\end{equation}
\noindent The superpotential has a definition

\begin{equation}
W(x)=-\left[\frac{d\ln\Psi_{0}^{(0)}(x)}{dx}\right].
\end{equation}

\noindent
Here, $\Psi_{0}^{(0)}(x)$ denotes the ground state wave
function that satisfies the following relation as

\beq
\Psi_{0}^{(0)}(x)=N_{0}~\exp\left[-\int^{x} W(x^{\prime})~dx^{\prime}\right].
\eeq
\smallskip
\noindent The partner Hamiltonians $\hat{H}_{1,~2}$ can also be
written in terms of the bosonic operators
$\hat{\Omega}^{-}$ and $\hat{\Omega}^{+}$

\begin{equation}
\hat{H}_{1}=\hat{\Omega}^{-}~\hat{\Omega}^{+}, \qquad
\hat{H}_{2}=\hat{\Omega}^{+}~\hat{\Omega}^{-},
\end{equation}

\noindent
where

\begin{equation}
\hat{\Omega}^{+}=\frac{d}{dx}+W(x),\qquad
\hat{\Omega}^{-}=-\frac{d}{dx}+W(x).
\end{equation}
\smallskip

\noindent It is remarkable to note that the energy eigenvalues of $\hat{H}_{1}$
and $\hat{H}_{2}$ are identical except for the ground state. In the case
of unbroken supersymmetry (SUSY), the ground state energy of the
Hamiltonian $\hat{H}_{1}$ is zero $\left(E_{0}^{(0)}=0\right)$ [11].
The straightforward procedure gives the first supersymmetric partner Hamiltonian $\hat{H}_{1}$ as
\smallskip
\begin{eqnarray}
\hat{H}_{1}(x)&=&-\frac{d^{2}}{dx^{2}}+V_{1}(x)\nonumber\\[0.2cm]
&=&(\hat{\Omega}_{1}^{-}~\hat{\Omega}_{1}^{+})+E_{1}^{(0)}.
\end{eqnarray}

\smallskip

\noindent Then, the Riccati equation for the superpotential $W_{1}\left(x\right)$ will take the form
\begin{equation}
W_{1}^{2}-W_{1}^{\prime}=V_{1}(x)-E_{1}^{(0)}.
\end{equation}

\smallskip
\noindent Now, we want to construct the second supersymmetric partner Hamiltonian
$\hat{H}_{2}$
\begin{eqnarray}
\hat{H}_{2}(x)&=&-\frac{d^{2}}{dx^{2}}+V_{2}(x)\nonumber\\[0.2cm]
        &=&\left(\hat{\Omega}_{2}^{+}~\hat{\Omega}_{2}^{-}\right)+E_{2}^{(0)}.
\end{eqnarray}

\noindent Therefore, the Riccati equation becomes
\begin{equation}
W_{2}^{2}+W_{2}^{\prime}=V_{2}(x)-E_{2}^{(0)}.
\end{equation}

\smallskip
\noindent By iterating the Hamiltonians $n-$th times similarly, one can get the
general Riccati equations as
\begin{eqnarray}
W_{n}^{2}-W_{n}^{\prime}=\left(\hat{\Omega}_{n}^{-}~\hat{\Omega}_{n}^{+}\right)+E_{n}^{(0)},\qquad
W_{n}^{2}+W_{n}^{\prime}=\left(\hat{\Omega}_{n}^{+}~\hat{\Omega}_{n}^{-}\right)+E_{n}^{(0)},
\end{eqnarray}

\noindent and
\begin{eqnarray}
\hat{H}_{n}(x)&=&-\frac{d^{2}}{dx^{2}}
+V_{n}(x)\nonumber\\[0.3cm]
        &=&\hat{\Omega}_{n}^{+}~\hat{\Omega}_{n}^{-}
+E_{n}^{(0)},\nonumber\\[0.3cm]
        &=&\hat{\Omega}_{n-1}^{-}~\hat{\Omega}_{n-1}^{+}
+E_{n-1}^{(0)},\quad\quad n=1,2,3,\ldots \end{eqnarray}

\noindent where
\smallskip
\begin{equation}
\hat{\Omega}_{n}^{+}=\frac{d}{dx}+\frac{d}
{dx}\left[\ln\Psi_{n}^{(0)}(x)\right],\qquad
\hat{\Omega}_{n}^{-}=-\frac{d}{dx}+\frac{d}
{dx}\left[\ln\Psi_{n}^{(0)}(x)\right].
\end{equation}


\smallskip

\noindent For unbroken SUSY case, the partner Hamiltonians satisfy
the energy eigenvalue and eigenfunction expressions as [11]

\begin{equation}
E_{n+1}^{(0)}=E_{n}^{(1)}, \qquad \quad E_{0}^{(0)}=0,\qquad
n=0,1,2,\ldots
\end{equation}

\smallskip
\noindent and

\begin{equation}
\Psi_{n}^{(0)}=\frac{1}{{\sqrt
{E_{n}^{(0)}}}}\left[\hat{\Omega}^{-}~\Psi_{n+1}^{(0)}\right],\quad\quad
\Psi_{n+1}^{(0)}=\frac{1}{{\sqrt
{E_{n}^{(0)}}}}\left[{\hat{\Omega}^{+}~\Psi_{n}^{(1)}}\right],
\end{equation}


\smallskip
\noindent with the same eigenvalue.

\section{Klein-Gordon Equation and Hulth\'{e}n Potential}
\noindent The Klein-Gordon (KG) equation is a relativistic version of the Schr\"{o}dinger equation.
In relativistic quantum mechanics, it describes the elementary particles with zero spin (spin-$0$)
quantum number (\emph{e. g.}, Higgs bosons). Its solutions implies a quantum scalar or pseudoscalar field.
There are various schemes on this equation by using different techniques. The $s-$wave
KG equation are studied for vector and scalar type of the
Hulth\'{e}n potential [37, 38]. The same problem are discussed for
scattering state solutions with the regular and irregular boundary
conditions [39]. For these potentials, the Green function are
obtained by means of the path integral approach [40]. Moreover, many
attempts have been made to develop approximation techniques such
as (shifted)large-N, $1/N$ expansion etc. for the KG equation with the
Coulomb like and Coulomb plus linear and Coulomb plus Aharonov-Bohm
potential [41-44]. Furthermore, the $s-$wave bound-state solutions
of the KG equation with the generalized Hulth\'{e}n potential are
examined by using the Nikiforov-Uvarov (NU) method [34-36], the
alternative SUSYQM approach [45] as well.\newline The $1-$D
Klein-Gordon equation for Lorentz vector and scalar potentials can
be defined as
\begin{equation}
\left\{-\frac{d^2}{dx^2}+[E-V(x)]^2-[m+S(x)]^2\right\}\Psi_{0}(x)=0,
\end{equation}
\noindent where $\Psi(x)=\frac{1}{x}\Psi_{0}(x)$, $V(x)$ and
$S(x)$ are Lorentz vector and scalar forms of the Hulth\'{e}n
potential.

\subsection{The Generalized Hulth\'{e}n Potential}
\noindent The \emph{q}-deformed Hulth\'{e}n potential is
\begin{equation}
V_{q}^{H}(x)=-V_{0}\frac{e^{-\lambda x}}{(1-q e^{-\lambda x})}.
\end{equation}

\noindent The Lorentz vector and scalar forms of the Hulth\'{e}n
potential can be written

\begin{equation}
V(x)=-V_{0}\frac{e^{-\lambda x}}{(1-q e^{-\lambda x})},\qquad
S(x)=-S_{0}\frac{e^{-\lambda x}}{(1-q e^{-\lambda x})}.
\end{equation}

\smallskip
\noindent Therefore, the effective Hulth\'{e}n potential becomes

\begin{equation}
V_{eff}(x)=[S^2(x)-V^2(x)]+2[m S(x)+E V(x)],
\end{equation}

\noindent $\epsilon=E^2-m^2<0$. We can write final form of the
effective potential as

\begin{equation}
V_{eff}(x)=\Gamma_{1}\frac{e^{-2\lambda x}}{(1-q e^{-\lambda
x})^{2}}-\Gamma_{2}\frac{e^{-\lambda x}}{(1-q e^{-\lambda x})},
\end{equation}

\bigskip
\noindent where $\Gamma_{1}=(S_0^2-V_0^2)>0$ and
$\Gamma_{2}=2(mS_0+EV_0)>0$.

\subsection{Supersymmetric Energy Bound Spectra of the KG Equation with Hulth\'{e}n Potential}

\noindent Now, we want to construct the successive superpotentials of the
\emph{q}-deformed Hulth\'{e}n potential, by proposing an ansatz
superpotential as

\begin{equation}
W_1(x)=-\nu_1\frac{e^{-\lambda x}}{(1-q e^{-\lambda x})}+\mu_1.
\end{equation}

\bigskip

\noindent The Riccati equation is defined in terms of  $ W_1(x)$
\begin{equation}
V_{eff}^{I}(x)=W_1^2(x)-W_1^{\prime}=V_{eff}(x)-\epsilon_0^{(1)}.
\end{equation}

\noindent Substituting the square of ansatz (21) and differential form of it into the equation (22) yields
\begin{eqnarray}
V_{eff}^{I}(x)&=& \nu_1(\nu_1-q\lambda)\frac{e^{-2\lambda x}}{(1-q
e^{-\lambda x})^{2}}-\nu_1(2\mu_1+\lambda)\frac{e^{-\lambda x}}{(1-q e^{-\lambda x})}+\mu_1^2\nonumber\\[0.4cm]
\widetilde{V}_{eff}(x)    &=&\Gamma_{1}\frac{e^{-2\lambda x}}{(1-q e^{-\lambda
x})^{2}}-\Gamma_{2}\frac{e^{-\lambda x}}{(1-q e^{-\lambda
x})}-\epsilon_0^{(1)}.
\end{eqnarray}

\noindent When we compare the right-sides of these equalities in the Eq. (23) term-by-term,

\begin{equation}
\mu_1=\frac{(\Gamma_1+q\Gamma_2-\nu_1^2)}{2q\nu_1}
\end{equation}

\noindent is obtained with $-\epsilon_0^{(1)}=\mu_1^2$. Now, we should determine
$V_{eff}^{I}(x)$ for the case
\begin{equation}
V_{eff}^{I}(x)=W_1^2(x)+W_1^{\prime}=V_{eff}(x)+{\mu_1^2}.
\end{equation}

\noindent Hence, one can write
\begin{equation}
V_{eff}^{I}(x)= \nu_1(\nu_1+q\lambda)\frac{e^{-2\lambda x}}{(1-q
e^{-\lambda x})^{2}}-\nu_1(2\mu_1-\lambda)\frac{e^{-\lambda x}}{(1-q
e^{-\lambda x})}+\mu_1^2.\\
\end{equation}

\smallskip

Let us propose the second ansatz superpotential for $W_2$ as

\begin{equation}
W_2(x)=-\nu_2\frac{e^{-\lambda x}}{(1-q e^{-\lambda x})}+\mu_2.
\end{equation}

\smallskip
\noindent Therefore, the Riccati equation will become
\begin{equation}
V_{eff}^{II}(x)=W_2^2(x)\mp W_2^{\prime}=V_{eff}(x)-\epsilon_0^{(2)}.
\end{equation}

\noindent By inserting the ansatz equation (27) into the equation (28), we obtain
\begin{eqnarray}
V_{eff}^{II}(x)&=& \nu_2(\nu_2\mp q\lambda)\frac{e^{-2\lambda x}}{(1-q
e^{-\lambda x})^{2}}-\nu_2(2\mu_2\pm\lambda)\frac{e^{-\lambda x}}{(1-q e^{-\lambda x})}+\mu_2^2\nonumber\\[0.3cm]
\overline{{V}}_{eff}(x) &=&\Gamma_{1}\frac{e^{-2\lambda x}}{(1-q e^{-\lambda
x})^{2}}-\Gamma_{2}\frac{e^{-\lambda x}}{(1-q e^{-\lambda
x})}-\epsilon_0^{(2)}.
\end{eqnarray}

\noindent From the Eqs. (23), (26) and (29), one gets

\begin{equation}
\nu_2\rightarrow\nu_1+q\lambda \qquad \textit{and} \qquad
\mu_2=\frac{(\Gamma_1+q\Gamma_2-\nu_2^2)}{2q\nu_2}.
\end{equation}

\noindent By using the relation $-\epsilon_0^{(2)}=\mu_2^2$,

\begin{equation}
\epsilon_0^{(2)}=-\left[\frac{\nu_2^2-(\Gamma_1+q\Gamma_2)}{2q\nu_2}\right]^2
\end{equation}

\noindent can be written.\\
\noindent Consequently, if the process is repeated $n$-times cautiously, we can
get the following results
\begin{eqnarray}
V_{eff}^{n}(x)&=& \nu_n(\nu_n+q\lambda)\frac{e^{-2\lambda x}}{(1-q
e^{-\lambda x})^{2}}-\nu_n(2\mu_n+\lambda)\frac{e^{-\lambda x}}{(1-q
e^{-\lambda x})}+\mu_n^2,\nonumber\\[0.3cm]
W_{n}(x)  &=&-\nu_n\frac{e^{-\lambda x}}{(1-q e^{-\lambda
x})}+\mu_n,\nonumber\\[0.4cm]
\epsilon_0^{(n)}&=&-\left[\frac{(\nu_n+nq\lambda)^2-(\Gamma_1+q\Gamma_2)}{2q(\nu_n+nq\lambda)}\right]^2; \qquad n=1, 2,\ldots
\end{eqnarray}

\noindent Use of notation $\epsilon_0^{(n)}=E_n^2-m^2$ gives the
energy eigenvalues of the general \emph{q}-deformed Hulth\'{e}n potential
for KG equation as
\smallskip

\begin{equation}
E_n=\pm\frac{i}{2q}\sqrt{\left[(\nu_n+nq\lambda)-\frac{(\Gamma_1+q\Gamma_2)}{(\nu_n+nq\lambda)}\right]^2-(4q^{2}m^2)}.
\end{equation}

\smallskip
\noindent Also, the $n$-th ground state wave function
can de defined from the Eq. (4)

\begin{equation}
\Psi_0^{(n)}=(1-q e^{\lambda x})^{\nu_n/\lambda}~e^{-\mu_n x}.
\end{equation}

\subsection{Bound Spectra of \textit{PT}-Supersymmetric Hulth\'{e}n Potential}
\noindent Let us now consider the \emph{PT}-symmetric form of the
Hulth\'{e}n potential by taking $\lambda\rightarrow i\lambda$ in the
Eq. (17). The effective potential can be written as follows

\begin{equation}
V_{eff}(x)=\Gamma_{1}\frac{e^{-2i\lambda x}}{(1-q e^{-i\lambda
x})^{2}}-\Gamma_{2}\frac{e^{-i\lambda x}}{(1-q e^{-i\lambda x})}.
\end{equation}

\smallskip
\noindent The ansatz superpotential of this potential will be
\begin{equation}
W_1(x)=-\nu_1\frac{e^{-i\lambda x}}{(1-q e^{-i\lambda x})}+\mu_1.
\end{equation}

\noindent Putting the ansatz equation (36) into the Eq. (22), we can obtain
\begin{eqnarray}
V_{eff}^{I}(x)&=& \nu_1(\nu_1-iq\lambda)\frac{e^{-2i\lambda x}}{(1-q
e^{-i\lambda x})^{2}}-\nu_1(2\mu_1+i\lambda)\frac{e^{-i\lambda x}}{(1-q e^{-i\lambda x})}+\mu_1^2\nonumber\\[0.4cm]
\widehat{{V}}_{eff}(x)&=&\Gamma_{1}\frac{e^{-2i\lambda x}}{(1-q e^{-i\lambda
x})^{2}}-\Gamma_{2}\frac{e^{-i\lambda x}}{(1-q e^{-i\lambda
x})}-\epsilon_0^{(1)}.
\end{eqnarray}

\noindent The same procedure followed in the above section leads to
the energy eigenvalues of KG equation for \emph{PT}-symmetric potential as

\begin{equation}
E_n=\pm\frac{i}{2q}\sqrt{\left[(\nu_n+inq\lambda)-\frac{(\Gamma_1+q\Gamma_2)}{(\nu_n+inq\lambda)}\right]^2-(4q^{2}m^2)}.
\end{equation}

\smallskip
\noindent The $n-$th ground state wave function
for the potential (36) can also be written

\begin{equation}
\Psi_0^{(n)}=(1-q e^{i\lambda x})^{\nu_n/i\lambda}~e^{-\mu_n x}.
\end{equation}

\subsection{Bound Spectra of \textit{PT}-Supersymmetric and Non-Hermitian Hulth\'{e}n Potential}
Let us assume the potential parameters $V_{0}\rightarrow (V_0+iV_I)$
and $\lambda\rightarrow i\lambda$ in Eq. (17). Then, the ansatz superpotential should become

\begin{equation}
W_1(x)=-\nu_1\frac{e^{-i\lambda x}}{(1-q e^{-i\lambda x})}+i\mu_1.
\end{equation}

\smallskip
\noindent By iterating the same procedure as in the section (3.2), we can obtain the energy
eigenvalues of the KG equation for the potential
\begin{equation}
E_n=\pm\frac{i}{2q}\sqrt{\left[(\nu_n+inq\lambda)-\frac{(\Gamma_1+q\Gamma_2)}{(\nu_n+inq\lambda)}\right]^2-(4q^{2}m^2)}.
\end{equation}

\noindent with $\epsilon_0^{(n)}<0$. Hence, the ground state wave function takes the form as

\begin{equation}
\Psi_0^{(n)}=(1-q e^{i\lambda x})^{\nu_n/i\lambda}~e^{-i\mu_n x}.
\end{equation}

\section{Concluding Remarks}

In this work, relativistic bound state spectra and the corresponding wave functions of
the Klein-Gordon (KG) equation with the generalized, \emph{PT}-symmetric,  \emph{PT}-supersymmetric and
non-Hermitian \emph{q}-deformed Hulth\'{e}n potential case are carried out by implementing
the Hamiltonian hierarchy procedure within the SUSYQM formalism. Various forms of the potential are also treated within the solution of the KG equation. We have proved that the bound state spectra of \emph{PT}-invariant
complex-valued non-Hermitian potentials may be real or complex depending on the actual potential parameters $(\emph{q}, \lambda, V_{0}, S_{0})$ in equations (17) and (18), and the superpotential parameters $(\nu_{1}, \mu_{1})$ in equation (21). For instance, in cases of $q=0$ referring to exponential potential, $q=1$ referring to the Hulth\'{e}n potential, $q=-1$ referring to the Woods-Saxon potential as well as with the condition of the Lorentz scalar and vector potentials $S_0=V_0$, the bound state results depend on the parameter $\Gamma_2=2(m+E)V_0$. Likewise, when $q=0$, the negative and positive bound energy results in equations (33), (38) and (41) will tend to go infinity automatically. Therefore, no explicit forms are existed in that case. However, in cases of $\emph{q}=\pm1$, there exist limitations on the bound state spectra. Furthermore, the corresponding ground state wave functions for some values of the potential parameters or only $x\rightarrow 0$ should explicitly require to be finite. As a final remark, achieving the results in the study are likely to leading new perspectives both in spontaneous and in dynamical supersymmetry breaking case studies as well as determining the optical spectra characteristics of particular semiconductor devices [62, 65].

\end{document}